\documentclass[aps,pra,reprint,groupedaddress,showpacs,showkeys,longbibliography]{revtex4-1}
\usepackage{amsmath}
\usepackage{graphicx}

\begin{document}

\title{Characterizing Feshbach resonances in ultracold scattering calculations}

\author{Matthew D. Frye}
\author{Jeremy M. Hutson}
\email{Author to whom correspondence should be addressed; j.m.hutson@durham.ac.uk}
\affiliation{Joint Quantum Centre (JQC) Durham-Newcastle, Department of
Chemistry, Durham University, South Road, Durham DH1 3LE, United Kingdom.}
\date{\today}

\begin{abstract}
We describe procedures for converging on and characterizing zero-energy
Feshbach resonances that appear in scattering lengths  for
ultracold atomic and molecular collisions as a function of an external field.
The \emph{elastic procedure} is appropriate for purely elastic scattering,
where the scattering length is real and displays a true pole. The
\emph{regularized scattering length} (RSL) procedure is appropriate when there
is weak background inelasticity, so that the scattering length is complex and
displays an oscillation rather than a pole, but the resonant scattering length
$a_{\textrm{res}}$ is close to real. The \emph{fully complex} procedure is
appropriate when there is substantial background inelasticity and the real and
imaginary parts of $a_{\textrm{res}}$ are required. We demonstrate these
procedures for scattering of ultracold $^{85}$Rb in various initial states. All
of them can converge on and provide full characterization of resonances, from
initial guesses many thousands of widths away, using scattering calculations at
only about 10 values of the external field.
\end{abstract}
\date{\today}
\maketitle

\section{Introduction}

Zero-energy Feshbach resonances are formed when a bound or quasi-bound state is
tuned across a threshold by varying an applied field, most commonly a magnetic
field. They are ubiquitous in studies of ultracold physics
\cite{Chin:RMP:2010}, where they can be used to tune scattering lengths for
many applications, including studies of equations of state
\cite{Nascimbene:2010}, solitons \cite{Frantzeskakis:2010}, and Efimov physics
\cite{Kraemer:2006, Huang:2nd-Efimov:2014}. They are also used for
magnetoassociation to form ultracold molecules \cite{Hutson:IRPC:2006,
Kohler:RMP:2006}.

Low-energy scattering may be described by the energy-dependent
scattering length $a(E,B)=-k^{-1}\tan\delta$, where $E=\hbar^2 k^2/2\mu$ is the
collision energy, $\mu$ is the reduced mass and $\delta$ is the scattering
phase shift. This is constant as $E\rightarrow 0$, where it reduces to the
usual zero-energy scattering length. At constant energy it is convenient to
write $a(E,B)$ as simply $a(B)$. In the simplest case of an isolated narrow
resonance without inelastic scattering, $a(B)$ is real and shows
a simple pole as a function of applied field $B$. If the background scattering
length $a_\mathrm{bg}(B)$ is constant across the width of the resonance, the
pole is described by \cite{Moerdijk:1995}
\begin{equation}
a(B)=a_\mathrm{bg} \left(1-\frac{\Delta}{B-B_\textrm{res}}\right),
\label{eq:pole}
\end{equation}
where $B_\textrm{res}$ is the position of the resonance, and the width of the
resonance is characterized by $\Delta$. The parameters are
generally weakly dependent on energy in the threshold region. Obtaining them
from quantum scattering calculations based on interaction potentials is an
important problem in ultracold collision physics.

It is possible to locate both the pole and the zero of the scattering length
and converge on them numerically using standard root-finding algorithms
\cite{Brue:LiYb:2012, Takekoshi:RbCs:2012}. In the case where
$a_\mathrm{bg}(B)$ is constant, $\Delta$ is the separation between the pole and
the zero. For resonances that are not isolated and narrow, the behavior of the
scattering length is more complicated than Eq.~\eqref{eq:pole}. Nevertheless,
Eq.~\eqref{eq:pole} always holds in some region close to the pole, and the
parameters may be defined in terms of this local behavior. When this is done,
$a_\mathrm{bg}$ may not describe $a(B)$ far from the pole and $\Delta$ may not
be precisely the separation between the pole and a zero. Such effects are
particularly prominent when there are numerous overlapping resonances
\cite{Jachymski:2013} or when $a_\mathrm{bg}$ is small, so that the zero is
artificially far from the pole \cite{Cho:RbCs:2013}.

If inelastic decay is present then the scattering length is complex
\cite{Balakrishnan:scat-len:1997} and its behavior is considerably more
complicated. It has no clearly defined zero-crossing, and it no longer shows a
pole but instead oscillates with a finite amplitude \cite{Hutson:res-note:2007,
Hutson:HeO2:2009}. This may render decayed resonances unsuitable for tuning
scattering lengths to large values \cite{Bohn:1997}. In addition, inelastic
rates usually peak sharply near resonance \cite{Gonzalez-Martinez:2007}, and
the resulting losses may make the resonances unsuitable for purposes such as
magnetoassociation \cite{Gonzalez-Martinez:LiYb:2013}. In other cases, Feshbach
resonances can actually \emph{reduce} inelastic cross sections, which might aid
sympathetic cooling \cite{Rowlands:2007, Hutson:HeO2:2009}.

In the inelastic case, there is no efficient procedure available
to locate and characterize Feshbach resonances. It is in principle possible to
obtain resonance parameters by explicit least-squares fitting of S-matrix
elements from quantum scattering calculations to appropriate functional forms
\cite{Ashton:1983}. It is also possible to extract an overall width by fitting
to the S-matrix eigenphase sum as a function of energy \cite{Ashton:1983}. This
approach has been used for zero-energy Feshbach resonances as a function of
magnetic field \cite{Gonzalez-Martinez:2007, Rowlands:2007}, but it requires
large numbers of scattering calculations and substantial manual labor. Better
methods are clearly needed.

In this paper we describe efficient, automatable procedures for
locating and characterizing zero-energy Feshbach resonances, both in the purely
elastic case and in the presence of inelastic scattering. Our algorithms are
built on an approach for resonances in purely elastic scattering that we have
used previously \cite{Cho:RbCs:2013, Zurn:Li2-binding:2013} but have not
described in detail. This converges towards a pole using an iterative 3-point
fit to calculated scattering lengths. We begin by describing an improved
algorithm for this case that converges stably on widths and background
scattering lengths as well as pole positions. We then extend the approach to
handle the important case when there is inelastic scattering but the inelastic
loss away from resonance is small. Finally we deal with the case where there is
strong background inelastic scattering. All the methods have been implemented
in the general-purpose quantum scattering package {\sc molscat}
\cite{molscat:2017}, and are illustrated here with examples from calculations
on collisions of $^{85}\textrm{Rb}$ \cite{Blackley:85Rb:2013}.

\section{Elastic scattering} \label{sec:el}

We first describe a reliable general method for converging on and
characterizing a resonance in the case of purely elastic scattering. Early
versions of this method have been employed in previous work
\cite{Cho:RbCs:2013, Zurn:Li2-binding:2013}, but here we refine it and provide
a complete description. We refer to the method described in this section as the
elastic procedure.

The elastic procedure uses three calculated scattering lengths $a_1$, $a_2$ and
$a_3$ at fields $B_1$, $B_2$ and $B_3$, respectively, close to the resonance.
Solving 3 simultaneous equations allows us to extract the three parameters from
Eq.~\eqref{eq:pole}. Defining
\begin{equation}
\rho = \left(\frac{B_3-B_1}{B_2-B_1}\right) \left(\frac{a_2-a_1}{a_3-a_1}\right),
\label{eq:rho}
\end{equation}
we obtain
\begin{align}
B_{\textrm{res}} &= \frac{B_3-B_2\rho}{1-\rho} \label{eq:bres} \\
a_{\textrm{bg}} \Delta &= \frac{(B_3-B_{\textrm{res}})(B_1-B_{\textrm{res}})(a_3-a_1)}{B_3-B_1}
\end{align}
and finally
\begin{equation}
a_{\textrm{bg}} = a_1 + \frac{a_{\textrm{bg}} \Delta}{B_1-B_{\textrm{res}}}.
\label{eq:abg}
\end{equation}

\begin{figure*}[tbp]
\centering
\includegraphics[height=0.85\textheight,clip=true,trim=2.9cm 3.4cm 2.8cm 1.8cm]{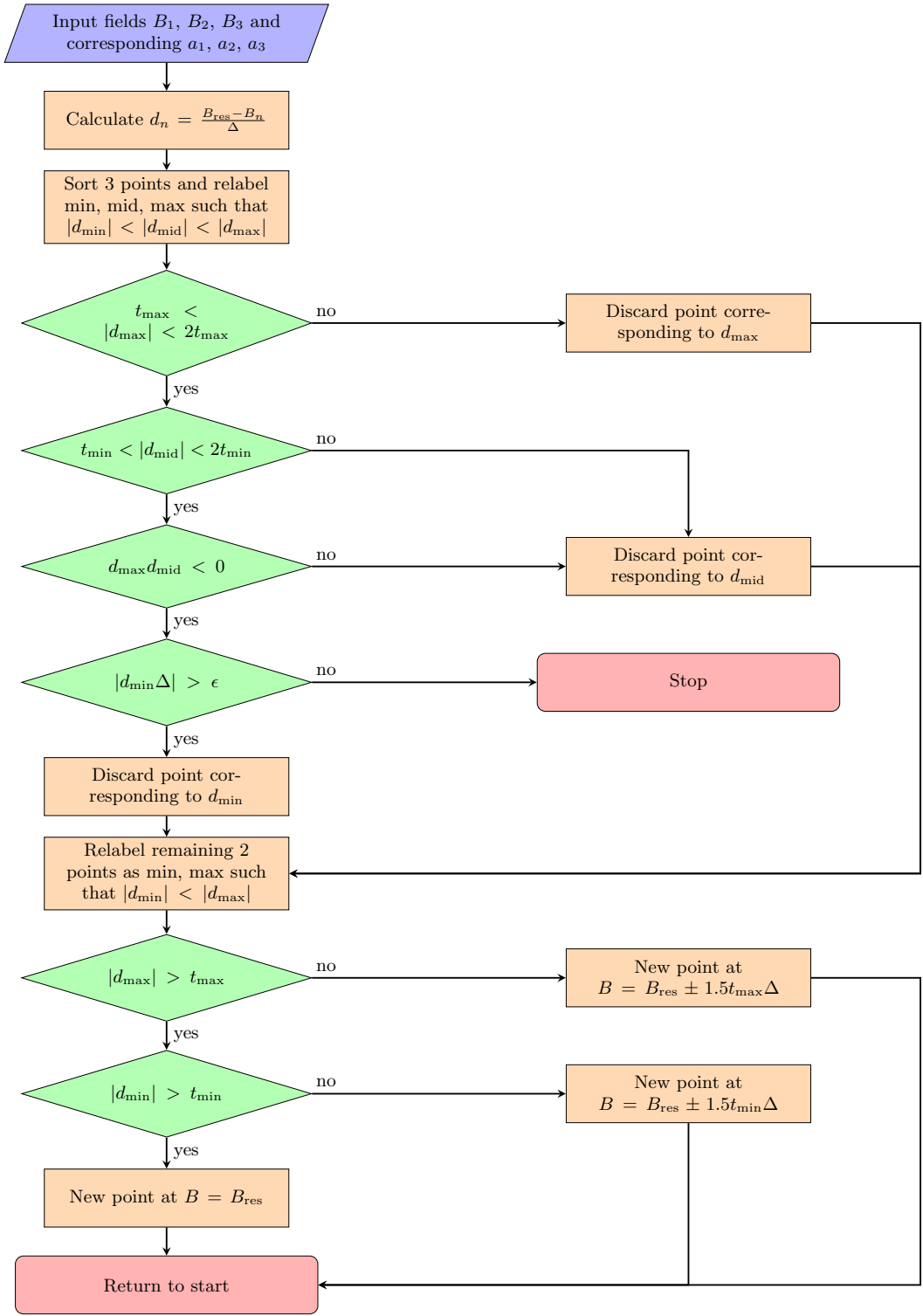}
\caption{Flowchart representation of the algorithm to select which point to
discard and where to place the next point. \label{fig:flowchart}}
\end{figure*}

In order to iterate and converge towards the pole we must not only choose a
point for a new scattering calculation but also choose which of the previous
three results to discard. The obvious choice for a new point is the estimated
$B_\textrm{res}$, but this causes points to pile up close to the pole, and
Eqs.~\eqref{eq:rho} to \eqref{eq:abg} are numerically unstable when 2 points
are very close together. We therefore choose the new point with the aim that
the final three points should include one point very close to the pole, one
point between $t_\textrm{min}\Delta$ and $2t_\textrm{min}\Delta$ from the pole,
and one point between $t_\textrm{max}\Delta$ and $2t_\textrm{max}\Delta$ from
the pole on the opposite side. These three points can be thought of as allowing
characterization of $B_{\textrm{res}}$, $a_{\textrm{bg}}\Delta$, and $a_{\rm
bg}$, respectively. The tolerances $t_\textrm{min}$ and
$t_\textrm{max}$ are positive, with $t_\textrm{min}<t_\textrm{max}$. The values
$t_\textrm{min}=0.1$ and $t_\textrm{max}=1.0$ are almost always appropriate for
isolated resonances; we use these values throughout this paper, but different
choices may be appropriate in other cases. We terminate the iteration when the
estimated value of $B_\textrm{res}$ is within a small amount $\epsilon$ of the
closest of the 3 points and the other two points satisfy the criteria above.
The logic we have implemented to select which point to discard and where to
place the next point is shown in Fig.\ \ref{fig:flowchart}.

We need 3 fields in the vicinity of the resonance to start the procedure. We
choose to use equally spaced points separated by a small amount $\delta B$; in
this work we choose this value to be 0.2~G. The algorithm will, of course,
perform best when one of the initial points is close to the pole, but in this
paper we choose points such that the pole is approximately at the midpoint of
two of them to provide the strictest test of the procedure. In practice, the
initial estimate of the pole position could come from a number of different
sources such as scattering calculations on a grid or calculations of the bound
states of the system; we usually use the program {\sc field}
\cite{Hutson:field:2011} which can directly calculate fields at which there is
a bound state exactly at threshold.

\begin{table*}
\caption{Convergence towards the resonance near 171~G for two $^{85}$Rb atoms
in their $F=2, M_F=2$ state. Units are G and the Bohr radius $a_0$.
\label{table:aa_b}}
\centering
\begin{tabular}{{c} | *{3}{c} | *{4}{c} }
\hline
 \multicolumn{8}{ c }{ Resonance near $B_\textrm{ref}=171.561$ G} \\  \hline
 & & & & \multicolumn{4}{| c }{ Estimated values} \\
 $n$	& $B_n-B_\textrm{ref}$	& $(B_n-B_\textrm{res})/\Delta$	& $a$				& $B_\textrm{res}-B_\textrm{ref}$	& $\Delta$		& $a_\textrm{bg}$	& $a_\textrm{bg}\Delta$\\ \hline
$1$ & $-1.00227\times 10^{-1}$ & $4.24\times 10^{3}$ & $-438.67$ & - & - & - & - \\
$2$ & $2.99773\times 10^{-1}$ & $-1.27\times 10^{4}$ & $-438.76$ & - & - & - & - \\
$3$ & $9.97730\times 10^{-2}$ & $-4.24\times 10^{3}$ & $-438.85$ & $3.40045\times 10^{-2}$ & $-1.8427\times 10^{-5}$ & $-438.73$ & $8.0846\times 10^{-3}$ \\
$4$ & $3.40045\times 10^{-2}$ & $-1.45\times 10^{3}$ & $-439.06$ & $3.68297\times 10^{-3}$ & $-2.0840\times 10^{-5}$ & $-438.75$ & $9.1437\times 10^{-3}$ \\
$5$ & $3.68297\times 10^{-3}$ & $-166$ & $-441.40$ & $-3.79856\times 10^{-4}$ & $-2.4633\times 10^{-5}$ & $-438.74$ & $1.0807\times 10^{-2}$ \\
$6$ & $-3.79856\times 10^{-4}$ & $6.49$ & $-371.13$ & $-2.26739\times 10^{-4}$ & $-2.3598\times 10^{-5}$ & $-438.75$ & $1.0354\times 10^{-2}$ \\
$7$ & $-2.26739\times 10^{-4}$ & $-0.00989$ & $-44657$ & $-2.26973\times 10^{-4}$ & $-2.3563\times 10^{-5}$ & $-438.76$ & $1.0339\times 10^{-2}$ \\
$8$ & $-2.23438\times 10^{-4}$ & $-0.150$ & $-3364.4$ & $-2.26973\times 10^{-4}$ & $-2.3568\times 10^{-5}$ & $-438.77$ & $1.0341\times 10^{-2}$ \\
$9$ & $-2.62324\times 10^{-4}$ & $1.50$ & $-146.31$ & $-2.26973\times 10^{-4}$ & $-2.3565\times 10^{-5}$ & $-438.82$ & $1.0341\times 10^{-2}$ \\
$10$ & $-2.26973\times 10^{-4}$ & $4.24\times 10^{-5}$ & $1.631\times 10^{7}$ & $-2.26972\times 10^{-4}$ & $-2.3564\times 10^{-5}$ & $-438.76$ & $1.0339\times 10^{-2}$ \\
\hline
\end{tabular}
\end{table*}

\begin{figure}
\includegraphics[width=1.00\columnwidth]{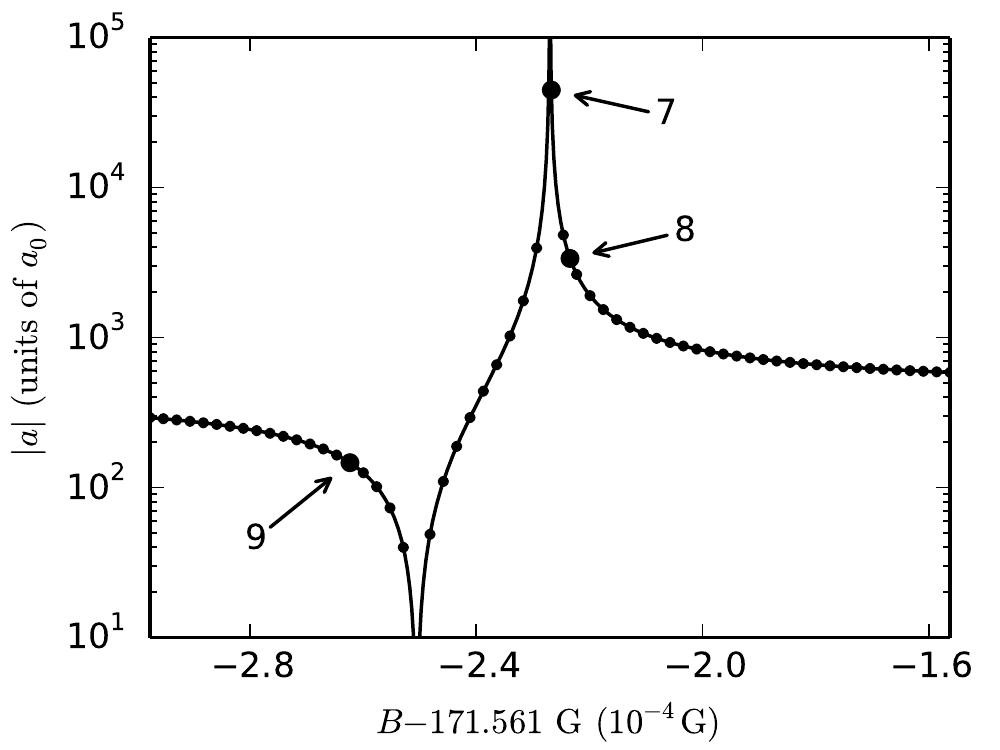}
\caption{Convergence towards the elastic resonance near 171~G for two $^{85}$Rb
atoms in their $F=2, M_F=2$ state. Note the logarithmic vertical scale. The
filled circles show the results of scattering calculations of $|a|$ at the
sequence of points $n$ produced by the elastic procedure; the black line shows
Eq.~\eqref{eq:pole} with the final estimated parameters; and the small dots
show scattering calculations on a grid for comparison. \label{fig:aa}}
\end{figure}

To demonstrate the convergence of this method, we apply it to a resonance near
171~G in collisions of two $^{85}$Rb atoms in their lowest ($F=2, M_F=2$)
state. Scattering lengths are calculated using the {\sc molscat} package, as
described by Blackley {\em et al.}\ \cite{Blackley:85Rb:2013}, at
energy $E=1\ \textrm{nK} \times k_\textrm{B}$. We choose $\epsilon=10^{-9}$~G,
which is limited by noise in our scattering calculations. Table
\ref{table:aa_b} summarizes the convergence towards the resonance, with the
parameters estimated by Eqs.~\eqref{eq:rho} to \eqref{eq:abg} at each
iteration; Figure \ref{fig:aa} provides a graphical representation of the
convergence process. This resonance is narrow, with $\Delta=2.3\times
10^{-5}$~G, yet our method successfully converges rapidly on the pole even
though the closest of the 3 initial points is over 4000 widths away. The 8th
and 9th points are actually placed away from the pole by the algorithm to
satisfy the requirements associated with $t_\textrm{min}$ and $t_\textrm{max}$
before the final point is placed extremely close to the pole. The entire
procedure needs only 10 scattering calculations and requires no human
intervention after the initial set of points; a corresponding manual search and
subsequent least-squares fit would have needed many more scattering
calculations and considerable human input.

If the pole position $B_{\textrm{res}}$ is all that is required, and $\Delta$
and $a_{\textrm{bg}}$ are unimportant, then the fastest convergence is often
achieved by setting $t_{\textrm{min}}=t_{\textrm{max}}=0$. With this choice,
the present algorithm reduces to that used in previous work from our group
\cite{Cho:RbCs:2013, Zurn:Li2-binding:2013}. The equations for $\Delta$ and
$a_{\textrm{bg}}$ then become unstable as convergence proceeds and the points
cluster close to the pole, but $B_{\textrm{res}}$ usually converges smoothly.

All the algorithms described here make the approximation that
$a_{\textrm{bg}}(B)$ is constant across the range of points. This approximation
improves as the convergence proceeds and the range of points becomes smaller.
Nevertheless, it is the limiting factor that determines the distance from which
convergence can be achieved. At least one of the initial points must give a
scattering length that is affected by the resonance by more than the variation
of $a_{\textrm{bg}}(B)$ across the range of the points. For very narrow
resonances, computational noise in the scattering length can also affect
convergence.

\section{Inelastic scattering}

In the presence of inelastic loss, the diagonal S-matrix element
in the incoming channel $S_{00}=\exp(2\textrm{i}\delta)$ has magnitude less
than 1. The phase shift $\delta$ is thus complex, and so is the scattering
length $a=\alpha- \textrm{i}\beta$, where $\beta\ge0$
\cite{Balakrishnan:scat-len:1997}. The real and imaginary parts of the
scattering length characterize the elastic and inelastic cross sections,
respectively. The energy-dependent scattering length may be written exactly as
\cite{Hutson:res-note:2007}
\begin{equation}
a(E,B)=\frac{-\tan\delta(E,B)}{k}=\frac{1}{\textrm{i}k}\left(\frac{1-S_{00}(E,B)}{1+S_{00}(E,B)}\right).
\end{equation}
Around a resonance, the scattering length at constant energy describes a circle
in the complex plane \cite{Hutson:res-note:2007}, beginning and ending at the
background scattering length $a_{\textrm{bg}}$,
\begin{equation}
a(B) = a_\textrm{bg} + \frac{a_\textrm{res}}{2(B-B_\textrm{res})/\Gamma_B^\textrm{inel} + \textrm{i}}.
\label{eq:decay_cplx}
\end{equation}
$a_\textrm{bg}=\alpha_\textrm{bg}-\textrm{i}\beta_\textrm{bg}$ is now complex
and $a_\textrm{res}=\alpha_\textrm{res}-\textrm{i}\beta_\textrm{res}$ is a
`resonant' scattering length that describes the size and direction of the
circle. $\Gamma_B^\textrm{inel}$ is a decay width for the quasibound state that
causes the resonance; it is a real quantity, with dimensions of field, whose
sign depends on the magnetic moment of the state relative to the threshold. It
is useful to identify
\begin{equation}
\alpha_\textrm{res}\Gamma_B^\textrm{inel}=-2 \alpha_\textrm{bg} \Delta
\label{eq:Gamma_Delta}
\end{equation}
to allow a connection back to Eq.~\eqref{eq:pole}, although $\Delta$ no longer
has a simple interpretation as the distance between the pole and zero in $a$.

Around a decayed resonance, both $\alpha$ and $\beta$ show an oscillation,
determined by $a_\textrm{res}$, rather than a pole \cite{Hutson:res-note:2007,
Rowlands:2007, Hutson:HeO2:2009}. This has implications for the observation and
use of such resonances \cite{Bohn:1997, Gonzalez-Martinez:2007,
Gonzalez-Martinez:LiYb:2013}. In the very common case $|a_{\textrm{res}}|\gg
\beta_{\textrm{bg}}$, $\beta(B)$ displays a peak of magnitude
$a_{\textrm{res}}$. However, $a_{\textrm{res}}$ is {\em inversely} proportional
to $\Gamma_B^\textrm{inel}$. Somewhat counterintuitively, therefore, {\em
weaker} inelastic decay of the quasibound state responsible for the resonance
causes a {\em higher} peak in $\beta(B)$ (and hence in the inelastic rate)
around $B_{\textrm{res}}$.

\subsection{Weak background inelasticity}

We first consider the important case where the background inelasticity can be
neglected, so we approximate $\beta_\textrm{bg}= 0$. Under this
approximation $a_\textrm{res}$ is also real \cite{Rowlands:2007}, though $a(B)$
itself remains complex near resonance. There are thus only 4 parameters to
extract. Even so, Eq.~\eqref{eq:decay_cplx} does not allow us to extract
parameters as easily as we could from Eq.~\eqref{eq:pole}. However, this can be
overcome by defining a `regularized scattering length'
\begin{align} \label{eq:a_regular}
\mathcal{A} &= \alpha + \frac{\beta^2}{\alpha-\alpha_\textrm{bg}} \\
&= \alpha_\textrm{bg}-\frac{\alpha_\textrm{bg}\Delta}{B-B_\textrm{res}}.
\end{align}
which is real and shows a simple pole just like Eq.\
\eqref{eq:pole}. This allows us to use Eqs.~\eqref{eq:rho} to \eqref{eq:abg}
with $a$ replaced by $\mathcal{A}$ to extract three of the parameters and
converge on the resonance position as before, with minimal modification of the
elastic procedure. We refer to the resulting method as the regularized
scattering length (RSL) procedure.

The final parameter $a_\textrm{res}$ can be estimated at each stage of the
convergence using the identity,
\begin{equation}
a_\textrm{res} = \frac{|a-a_\textrm{bg}|^2}{\beta} = \beta +
\frac{(\alpha-\alpha_\textrm{bg})^2}{\beta}.
\label{eq:a_res}
\end{equation}
In the important case where $\Gamma_B^\textrm{inel}$ is very small, the peak in
$\beta$ is very narrow. Estimating $a_\textrm{res}$ from the maximum value of
$\beta$ can thus be very difficult, but Eq.~\eqref{eq:a_res} provides a useful
estimate as soon as both $\alpha$ and $\beta$ differ significantly from their
background values. Equations \eqref{eq:a_regular} and \eqref{eq:a_res} each
need an estimate of $\alpha_\textrm{bg}$. This can be obtained iteratively, but
we find that in practice it is adequate to take it from the previous or current
iteration, respectively. To calculate $\mathcal{A}$ at the first iteration we
use the average of $a_1$ and $a_2$ as an initial approximation for
$a_\textrm{bg}$. Equation \eqref{eq:a_res} can also be used separately from the
convergence algorithm employed here, for example to estimate $a_\textrm{res}$
from scattering calculations on a grid that is not fine enough to resolve the
peak in $\beta$.

\begin{table*}[tbhp]
\caption{Convergence towards resonances with weak background inelasticity for
two $^{85}$Rb atoms in their $F=2, M_F=-2$ state. Units are G and the Bohr
radius $a_0$. \label{table:ee}}
\centering
\begin{tabular}{ c | *{5}{c} | *{4}{c} }
\hline
\multicolumn{10}{ c }{ Resonance near $B_\textrm{ref}=215.084$ G} \\  \hline
 & & & & & & \multicolumn{4}{| c }{ Estimated values} \\
  $n$	& $B_n-B_\textrm{ref}$	& $(B_n-B_\textrm{res})/\Delta$	& $\alpha_n$	& $\beta_n$		& $\mathcal{A}_n$	& $B_\textrm{res}-B_\textrm{ref}$	& $\Delta$		& $a_\textrm{bg}$	& $a_\textrm{res}$	\\ \hline
$1$ & $-9.96246\times 10^{-2}$ & $-18.0$ & $-402.1$ & $0.000796$ & $-402.1$ & - & - & - & - \\
$2$ & $3.00375\times 10^{-1}$ & $53.9$ & $-374.5$ & $0.000692$ & $-374.5$ & - & - & - & - \\
$3$ & $1.00375\times 10^{-1}$ & $18.0$ & $-360.0$ & $0.000649$ & $-360.0$ & $2.81284\times 10^{-3}$ & $5.514\times 10^{-3}$ & $-381.53$ & $7.168\times 10^{5}$ \\
$4$ & $2.81284\times 10^{-3}$ & $0.438$ & $489.6$ & $0.00219$ & $489.6$ & $3.95129\times 10^{-4}$ & $5.524\times 10^{-3}$ & $-381.02$ & $3.459\times 10^{8}$ \\
$5$ & $3.95129\times 10^{-4}$ & $0.00354$ & $1.07\times 10^{5}$ & $67.6$ & $1.07\times 10^{5}$ & $3.75429\times 10^{-4}$ & $5.568\times 10^{-3}$ & $-381.19$ & $1.716\times 10^{8}$ \\
$6$ & $-7.97622\times 10^{-3}$ & $-1.50$ & $-635.1$ & $0.00197$ & $-635.1$ & $3.75433\times 10^{-4}$ & $5.569\times 10^{-3}$ & $-381.01$ & $1.716\times 10^{8}$ \\
$7$ & $1.21083\times 10^{-3}$ & $0.150$ & $2159$ & $0.0299$ & $2159$ & $3.75433\times 10^{-4}$ & $5.569\times 10^{-3}$ & $-381.00$ & $1.716\times 10^{8}$ \\
$8$ & $3.75433\times 10^{-4}$ & $-1.80\times 10^{-7}$ & $-1.32\times 10^{7}$ & $1.70\times 10^{8}$ & $-2.19\times 10^{9}$ & $3.75434\times 10^{-4}$ & $5.569\times 10^{-3}$ & $-381.00$ & $1.707\times 10^{8}$ \\
\hline \hline
 \multicolumn{10}{ c }{ Resonance near $B_\textrm{ref}=603.977$ G} \\ \hline
$1$ & $-9.93851\times 10^{-2}$ & $-531$ & $-476.7$ & $0.00159$ & $-476.7$ & - & - & - & - \\
$2$ & $3.00615\times 10^{-1}$ & $1.59\times 10^{3}$ & $-475.6$ & $1.39\times 10^{-5}$ & $-475.6$ & - & - & - & - \\
$3$ & $1.00615\times 10^{-1}$ & $531$ & $-475.0$ & $0.000630$ & $-475.0$ & $1.08784\times 10^{-2}$ & $1.7996\times 10^{-4}$ & $-475.91$ & $1.446\times 10^{3}$ \\
$4$ & $1.08784\times 10^{-2}$ & $54.5$ & $-467.1$ & $0.0954$ & $-467.1$ & $9.67212\times 10^{-4}$ & $1.8191\times 10^{-4}$ & $-475.82$ & $7.991\times 10^{2}$ \\
$5$ & $9.67212\times 10^{-4}$ & $1.87$ & $-246.8$ & $76.3$ & $-221.4$ & $6.13682\times 10^{-4}$ & $1.8905\times 10^{-4}$ & $-475.86$ & $7.635\times 10^{2}$ \\
$6$ & $6.13682\times 10^{-4}$ & $-0.00659$ & $-483.8$ & $762$ & $-7.32\times 10^{4}$ & $6.14914\times 10^{-4}$ & $1.8838\times 10^{-4}$ & $-475.82$ & $7.621\times 10^{2}$ \\
$7$ & $5.86657\times 10^{-4}$ & $-0.150$ & $-648.9$ & $720$ & $-3647$ & $6.14919\times 10^{-4}$ & $1.8837\times 10^{-4}$ & $-475.81$ & $7.621\times 10^{2}$ \\
$8$ & $6.14919\times 10^{-4}$ & $-2.65\times 10^{-5}$ & $-475.8$ & $762$ & $-1.84\times 10^{7}$ & $6.14924\times 10^{-4}$ & $1.8838\times 10^{-4}$ & $-475.83$ & $7.621\times 10^{2}$ \\
\hline
\end{tabular}
\end{table*}

\begin{figure}
\includegraphics[width=0.95\columnwidth]{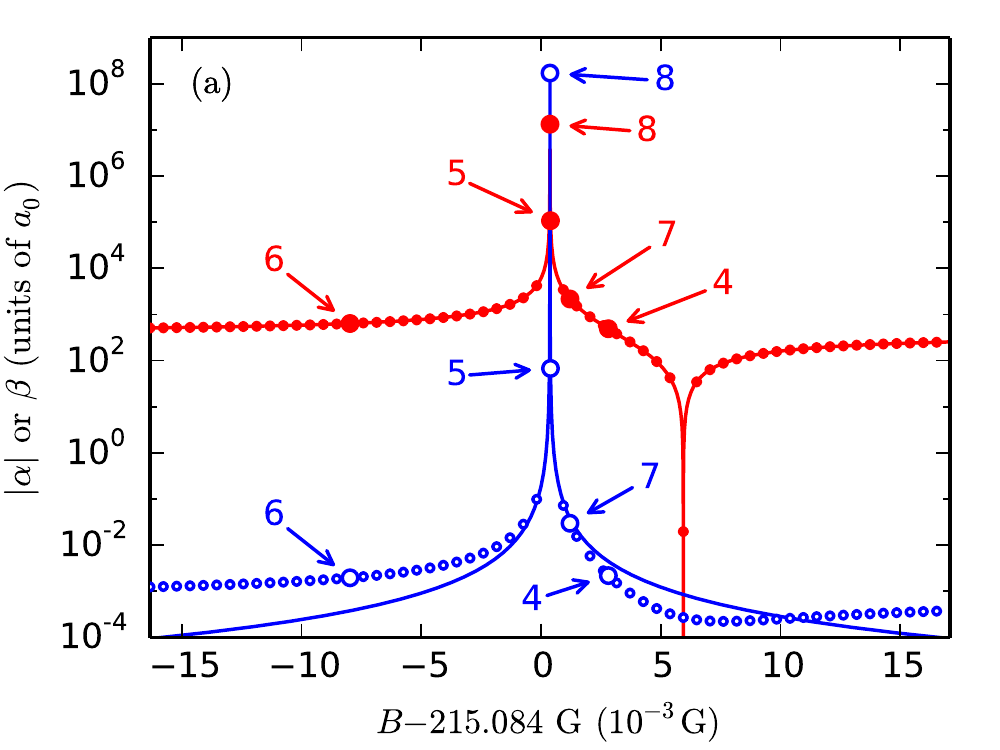}
\includegraphics[width=0.95\columnwidth]{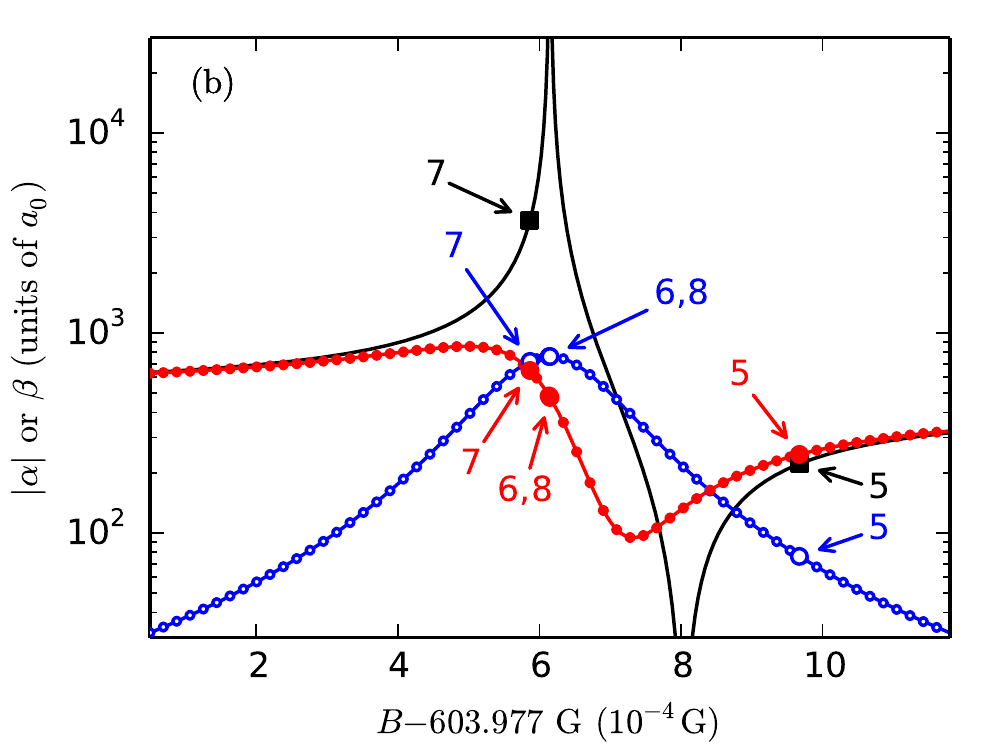}
\caption{Convergence towards resonances with weak background inelasticity for
two $^{85}$Rb atoms in their $F=2, M_F=-2$ state. (a) Resonance near 215~G; (b)
Resonance near 604~G. The large symbols show values of $|\alpha|$ (red, filled
circles) and $\beta$ (blue, open circles) from scattering calculations at the
sequence of points $n$ produced by the RSL procedure; the fitted functions are
shown as corresponding lines through the symbols; the small dots show
scattering calculations on a grid for comparison. The black squares and line
in (b) show the regularized scattering length $\mathcal{A}$. \label{fig:ee}}
\end{figure}

Table \ref{table:ee} summarizes the convergence towards two resonances in
collisions of a pair of $^{85}$Rb atoms in their $F=2, M_F=-2$ excited state,
using the RSL procedure. These results are also shown in Fig.\ \ref{fig:ee}.
These collisions are weakly inelastic away from resonances, because loss comes
only from spin-relaxation transitions driven by the weak dipole-dipole
interaction. We use a slightly larger value for the convergence criterion than
in the previous section, $\epsilon = 10^{-8}$~G.

The first inelastic resonance we analyze, near 215~G, shows only weak inelastic
decay, as seen from the small values of $\beta$ and negligible differences
between $\alpha$ and $\mathcal{A}$ except at the final point. The RSL procedure
converges smoothly and provides stable values of all the resonance parameters.
The fitted $\beta(B)$ is shown in Fig.\ \ref{fig:ee}(a); it is accurate near
the center of the resonance, but deviates from the calculated values by a small
amount in the wings because the actual background $\beta_\textrm{bg}$ is
non-zero. As described above, the RSL procedure provides an estimate
$a_\textrm{res}=1.7\times10^8$~$a_0$ that is stable over the final few
iterations even when $\beta$ is 6 orders of magnitude smaller than
$a_\textrm{res}$; the final calculation confirms that these estimates of
$a_\textrm{res}$ are remarkably accurate. For this resonance, the elastic
procedure would work well until the last point, when it would predict a pole
position some distance away from the resonance. The elastic procedure would
thus fail to converge, and continue indefinitely, repeatedly approaching the
resonance and jumping away again.

The second resonance we analyze, near 604~G, is quite strongly decayed. The
pole in $\alpha$ is strongly suppressed, to the point that $\alpha$ does not
even cross zero. By contrast, the regularized scattering length still has a
pole and zero crossing as before. The elastic procedure would fail completely
anywhere near the center of the resonance, but with the modification of Eq.\
\eqref{eq:a_regular} we can efficiently converge to the resonance position. The
final fitted $\alpha(B)$ and $\beta(B)$, shown in Fig.\ \ref{fig:ee}(b), agree
very well with the calculated values, demonstrating that the resonance has been
accurately characterized. The new fitted value of $\Delta=1.8\times10^{-4}$~G
is two orders of magnitude smaller than the value reported previously
\cite{Blackley:85Rb:2013}, which was obtained by fitting $\alpha(B)$ to Eq.\
\eqref{eq:pole} far from resonance.

\subsection{Strong background inelasticity}

Finally, we consider the case with background inelasticity
included. There are now a total of 6 parameters required to characterize a
resonance according to Eq.~\eqref{eq:decay_cplx}: $B_\textrm{res}$,
$\Gamma_B^\textrm{inel}$, and the real and imaginary parts of $a_\textrm{bg}$
and $a_\textrm{res}$. However, each value of $a(B)$ has real and imaginary
parts, so we again need scattering calculations at only three fields.

We begin by locating the scattering length at the center of the circle
described by Eq.~\eqref{eq:decay_cplx},
$a_\textrm{c}=a_\textrm{bg}-\textrm{i}a_\textrm{res}/2$. Starting from the
equation for a circle,
$(\alpha_n-\alpha_\textrm{c})^2+(\beta_n-\beta_\textrm{c})^2=R^2$, it is
straightforward to derive the simultaneous equations
\begin{equation}
\begin{pmatrix} \alpha_2-\alpha_1 & \beta_2-\beta_1 \\ \alpha_3-\alpha_2 & \beta_3-\beta_2 \end{pmatrix}
\begin{pmatrix} \alpha_\textrm{c} \\ \beta_\textrm{c} \end{pmatrix}
= \frac{1}{2} \begin{pmatrix} |a_2|^2-|a_1|^2 \\ |a_3|^2-|a_2|^2  \end{pmatrix}.
\end{equation}
These are solved to obtain $a_\textrm{c}$ and $R=|a_n-a_\textrm{c}| =
|a_\textrm{res}|/2$. Across the resonance, the angle $\theta$ around this
circle is described by a Breit-Wigner phase,
\begin{equation}
\frac{\theta}{2}=\frac{\theta_{\textrm{bg}}}{2} + \arctan\left(\frac{\Gamma^{\rm
inel}}{2(B_{\textrm{res}}-B)}\right).
\end{equation}
We define the dimensionless quantity
\begin{equation}
\widetilde{a}(B)=\tan\frac{\theta}{2}=\tan\left(\frac{\arg (a(B)-a_\textrm{c})}{2}\right),
\end{equation}
which has a pole analogous to Eq.\ \eqref{eq:pole}. We evaluate
$\widetilde{a}_1$, $\widetilde{a}_2$ and $\widetilde{a}_3$ at $B_1$, $B_2$ and
$B_3$ and use Eqs.~\eqref{eq:rho} to \eqref{eq:abg} to obtain parameters
$\widetilde{B}_\textrm{res}$, $\widetilde{\Delta}$, and
$\widetilde{a}_\textrm{bg}$ (which do not have immediate physical
interpretations).
$\widetilde{a}_\textrm{bg}=\tan(\theta_\textrm{bg}/2)$ tells us
where on the circle $a_\textrm{bg}$ lies,
\begin{equation}
a_\textrm{bg}=a_\textrm{c}+R\exp(\textrm{i}\theta_\textrm{bg})
\end{equation}
and therefore
\begin{equation}
a_\textrm{res}=2\textrm{i}(a_\textrm{c}-a_\textrm{bg}).
\end{equation}
$a(B_{\textrm{res}})$ is diametrically opposite $a_\textrm{bg}$ on the circle,
so
\begin{equation}
\widetilde{a}(B_{\textrm{res}})=\tan\left(\frac{\theta_\textrm{bg}+\pi}{2}\right)
=-\frac{1}{\widetilde{a}_\textrm{bg}}.
\end{equation}
We then obtain $B_{\textrm{res}}$ from
\begin{equation}
B_\textrm{res}=\widetilde{B}_\textrm{res}-\frac{\widetilde{a}_\textrm{bg}
\widetilde{\Delta}}{\widetilde{a}(B_{\textrm{res}})-\widetilde{a}_\textrm{bg}} =
\widetilde{B}_\textrm{res}+\frac{\widetilde{\Delta}}{1+\widetilde{a}_\textrm{bg}^{-2}}.
\end{equation}
Finally, we obtain $\Gamma_B^\textrm{inel}$ from one calculated scattering
length using Eq.\ \eqref{eq:decay_cplx}.

This procedure provides an estimate of $B_{\textrm{res}}$ and other parameters
from calculations of $a(B)$ at a set of 3 points. We iterate using the
algorithm described in section \ref{sec:el}, but using the larger of
$\Gamma_B^\textrm{inel}$ and $\Delta$ to constrain the separation of the points
from $B_{\textrm{res}}$. We refer to the resulting method as the fully complex
procedure.

\begin{figure}
\includegraphics[width=0.95\columnwidth]{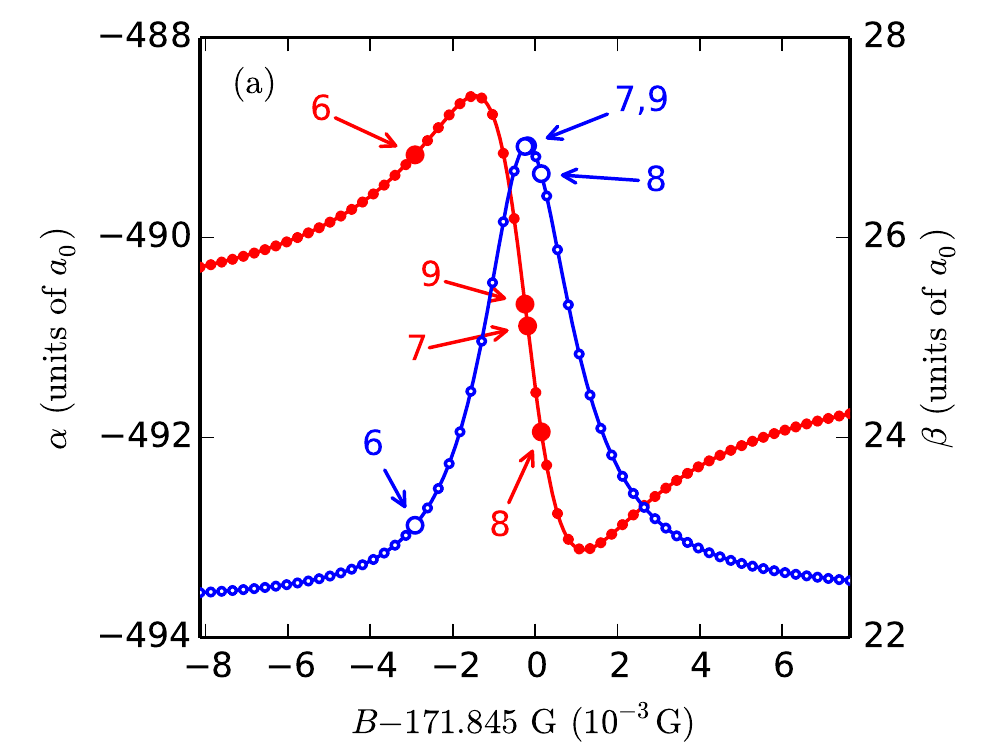}
\includegraphics[width=0.95\columnwidth]{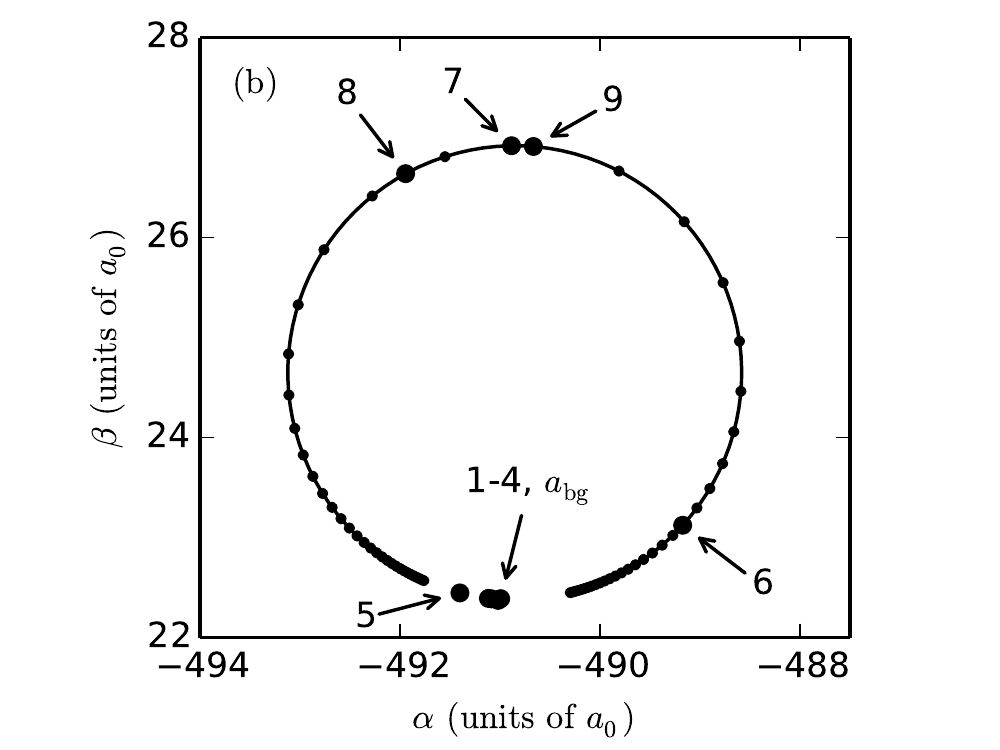}
\caption{Convergence towards the resonance near 172 G for two $^{85}$Rb atoms
in their $F=3, M_F=2$ state. (a) $\alpha$ (red, filled circles) and $\beta$
(blue, open circles) from scattering calculations at the sequence of points $n$
produced by the fully complex procedure; the fitted functions are shown as
corresponding lines through the symbols; the small dots show scattering
calculations on a grid for comparison. (b) The circle described by
$a(B)=\alpha(B)-\textrm{i}\beta(B)$ in the complex plane. \label{fig:kk}}
\end{figure}

To demonstrate this, we consider convergence towards a resonance near 172~G in
collisions of two $^{85}$Rb atoms in their $F=3, M_F=2$ excited state. In this
case the atoms can decay through spin-exchange collisions, which cause faster
inelastic loss away from resonance than in Sec.\ III A. The convergence is
summarized in Table \ref{table:kk_a} and shown in Fig.\ \ref{fig:kk},
using $\epsilon = 10^{-7}$~G. The procedure converges rapidly on
the resonance position and the final fitted functions show excellent agreement
with the calculated scattering lengths. The resonance is very strongly decayed;
$|a_\textrm{res}|$ is less than $5\ a_0$ and has a substantial imaginary
component. This makes the oscillations in $\alpha(B)$ and $\beta(B)$ somewhat
asymmetric.

\begin{table*}
\caption{Convergence towards the resonance near 172 G for two $^{85}$Rb atoms
in their $F=3, M_F=2$ state. Units are G and the Bohr radius $a_0$.
\label{table:kk_a}}
\centering
\begin{tabular}{ c | *{4}{c} | *{6}{c} }
 \hline
 \multicolumn{11}{ c }{ Resonance near $B_\textrm{ref}=171.845$ G} \\  \hline
 & & & & & \multicolumn{6}{| c }{ Estimated values} \\
 $n$	& $B_n-B_\textrm{ref}$	& $(B_n-B_\textrm{res})/\Delta$	& $\alpha_n$	& $\beta_n$	& $B_\textrm{res}-B_\textrm{ref}$	& $\Gamma_B^\textrm{inel}$		& $\alpha_\textrm{bg}$	& $\beta_\textrm{bg}$	& $\alpha_\textrm{res}$	& $\beta_\textrm{res}$	\\ \hline
$1$ & $-1.00244\times 10^{-1}$ & $38.0$ & $-490.99$ & $22.388$ & - & - & - & - & - & - \\
$2$ & $2.99756\times 10^{-1}$ & $-114$ & $-491.02$ & $22.371$ & - & - & - & - & - & - \\
$3$ & $9.97560\times 10^{-2}$ & $-38.0$ & $-491.09$ & $22.386$ & $6.92055\times 10^{-2}$ & $-4.7788\times 10^{-2}$ & $-491.01$ & $22.377$ & $0.10979$ & $0.065852$ \\
$4$ & $6.92055\times 10^{-2}$ & $-26.4$ & $-491.12$ & $22.391$ & $1.58950\times 10^{-2}$ & $-1.2632\times 10^{-2}$ & $-491.03$ & $22.384$ & $0.71330$ & $0.026746$ \\
$5$ & $1.58950\times 10^{-2}$ & $-6.14$ & $-491.40$ & $22.446$ & $-2.91246\times 10^{-3}$ & $-1.5078\times 10^{-3}$ & $-491.01$ & $22.376$ & $9.7150$ & $-1.3474$ \\
$6$ & $-2.91246\times 10^{-3}$ & $1.01$ & $-489.17$ & $23.122$ & $-1.80823\times 10^{-4}$ & $-2.7312\times 10^{-3}$ & $-491.03$ & $22.382$ & $4.4502$ & $-0.37429$ \\
$7$ & $-1.80823\times 10^{-4}$ & $-0.0241$ & $-490.88$ & $26.918$ & $-2.43111\times 10^{-4}$ & $-2.6270\times 10^{-3}$ & $-491.04$ & $22.386$ & $4.5243$ & $-0.36788$ \\
$8$ & $1.50937\times 10^{-4}$ & $-0.150$ & $-491.94$ & $26.638$ & $-2.44221\times 10^{-4}$ & $-2.6291\times 10^{-3}$ & $-491.04$ & $22.387$ & $4.5232$ & $-0.37363$ \\
$9$ & $-2.44221\times 10^{-4}$ & $1.75\times 10^{-6}$ & $-490.67$ & $26.910$ &  $-2.44216\times 10^{-4}$ & $-2.6290\times 10^{-3}$ & $-491.04$ & $22.387$ & $4.5232$ & $-0.37361$ \\
\hline
\end{tabular}
\end{table*}

\begin{figure}
\includegraphics[width=0.95\columnwidth]{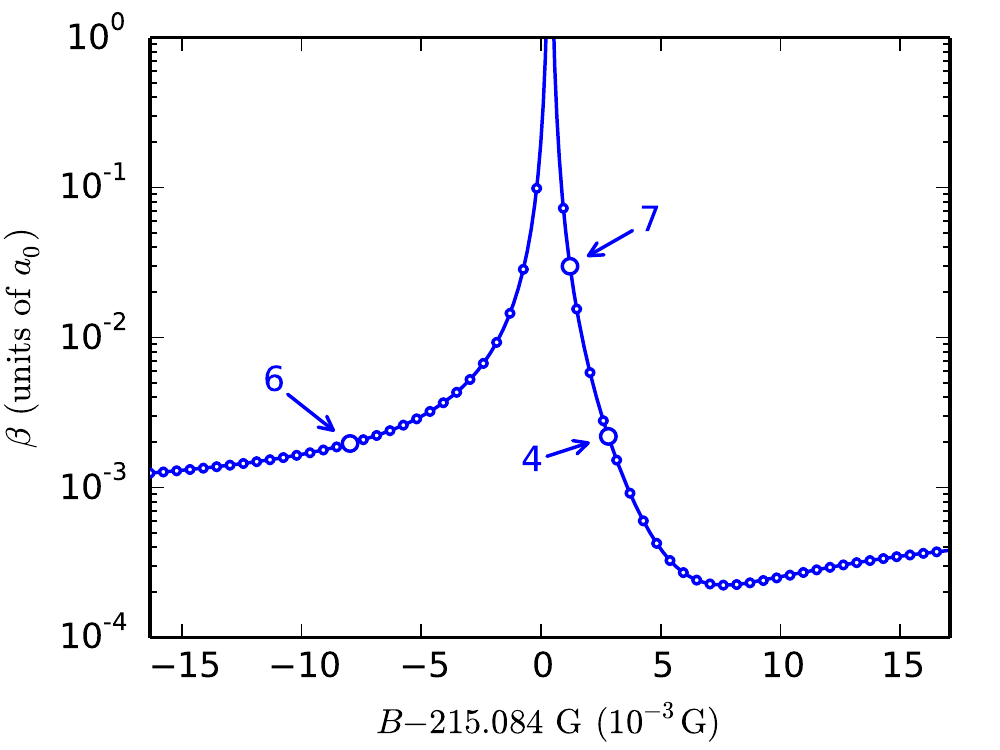}
\caption{Convergence towards the resonance near 215~G for two $^{85}$Rb atoms
in their $F=2, M_F=-2$ state using the fully complex procedure. Only
$\beta$ is shown and the axis is expanded to show the asymmetry clearly.
Symbols and lines are as in previous figures. \label{fig:ee_b_full}}
\end{figure}

The fully complex procedure can also resolve the discrepancy between the
calculated $\beta(B)$ and the fitted function far from resonance in Fig.\
\ref{fig:ee}(a). Figure \ref{fig:ee_b_full} shows the results of the fully
complex procedure in this case, and it may be seen that excellent agreement is
obtained. The converged values of the parameters are very similar to those in
Table \ref{table:ee}, with the addition of
$\beta_{\textrm{bg}}=7.20\times10^{-4}\ a_0$ and $\beta_{\textrm{res}}=-582\
a_0$.

For this procedure to converge well, the circle in the complex plane described
by $a(B)$ must be well formed. Variation of $a_\textrm{bg}(B)$ across the width
of the resonance can distort the circle; if this distortion is significant
compared to the size of the circle, the procedure may fail. This leads to the
criterion
\begin{equation}
\left|\frac{da_\textrm{bg}}{dB} \Gamma_B^\textrm{inel} \right| \ll |a_\textrm{res}|.
\label{eq:criterion}
\end{equation}
The procedure may thus be unsuitable for the widest and most strongly decayed
resonances (large $\Gamma_B^\textrm{inel}$ and small $a_\textrm{res}$). The
procedure may also fail for overlapping resonances. These restrictions are
similar to the criteria used to define an isolated narrow resonance
\cite{Ashton:1981, Ashton:1983}.

\section{Conclusions}

In this paper we have developed three procedures for efficiently and accurately
converging on and characterizing different kinds of zero-energy Feshbach
resonances as a function of external field. These procedures can converge on
and accurately characterize resonances, from initial guesses many thousands of
widths away, with a total of only around 10 scattering calculations.

First we have described the \emph{elastic procedure}. This is designed for
resonances in purely elastic scattering, where the scattering length has a true
pole. At each iteration, the procedure characterizes the resonance using
scattering calculations at 3 values of the external field, while ensuring that
the points do not cluster too close to the pole. This allows stable evaluation
of the width and background scattering length as well as the pole position.

For the case of weak background inelasticity we have developed the
\emph{regularized scattering length} (RSL) procedure. The oscillation in the
complex scattering length is converted into a true pole in a ``regularized''
scattering length, and convergence on the pole is achieved in the same way as
in the elastic procedure. We also provide a means to estimate the resonant
scattering length $a_\textrm{res}$ from calculations in the wings of the
resonance.

Finally, we have developed a \emph{fully complex procedure} to converge on and
extract all 6 parameters needed to characterize resonances when there is
substantial background inelasticity and the real and imaginary parts of
$a_{\textrm{res}}$ are required.

\acknowledgments

The authors are grateful to C. Ruth Le Sueur for valuable discussions on
implementation of these procedures in the {\sc molscat} program. This work has
been supported by the UK Engineering and Physical Sciences Research Council
(grant EP/I012044/1).

\bibliography{../all,Decayed_Resonance}
\end{document}